\documentclass[aip,twocolumn,apl,reprint,amsmath,amssymb,superscriptaddress]{revtex4-1}

\usepackage{graphicx}
\usepackage{dcolumn}
\usepackage{bm}

\begin{document}

\title{Graphene microwave transistors on sapphire substrates}

\author{E. Pallecchi}\email[These authors contribute equally to this work]{}
\affiliation{Laboratoire Pierre Aigrain, Ecole Normale Sup\'erieure, CNRS (UMR 8551),
Universit\'e P. et M. Curie, Universit\'e D. Diderot,
24, rue Lhomond, 75231 Paris Cedex 05, France}

\author{C. Benz}\email[These authors contribute equally to this work]{}
\affiliation{Institute of Nanotechnology, Karlsruhe Institute of Technology, D-76021 Karlsruhe, Germany}
\affiliation{Institute of Physics, Karlsruhe Institute of Technology, D-76049 Karlsruhe, Germany}

\author{A.C. Betz}
\affiliation{Laboratoire Pierre Aigrain, Ecole Normale Sup\'erieure, CNRS (UMR 8551),
Universit\'e P. et M. Curie, Universit\'e D. Diderot,
24, rue Lhomond, 75231 Paris Cedex 05, France}

\author{H.v. L\"{o}hneysen}
\affiliation{Institute of Nanotechnology, Karlsruhe Institute of Technology, D-76021 Karlsruhe, Germany}
\affiliation{Institute of Physics, Karlsruhe Institute of Technology, D-76049 Karlsruhe, Germany}
\affiliation{Institute for Solid State Physics, Karlsruhe Institute of Technology, D-76021 Karlsruhe, Germany}

\author{B. Pla\c{c}ais}
\affiliation{Laboratoire Pierre Aigrain, Ecole Normale Sup\'erieure, CNRS (UMR 8551),
Universit\'e P. et M. Curie, Universit\'e D. Diderot,
24, rue Lhomond, 75231 Paris Cedex 05, France}

\author{R. Danneau}\email[Author to whom correspondence should be addressed. Electronic mail: romain.danneau@kit.edu]{}
\affiliation{Institute of Nanotechnology, Karlsruhe Institute of Technology, D-76021 Karlsruhe, Germany}
\affiliation{Institute of Physics, Karlsruhe Institute of Technology, D-76049 Karlsruhe, Germany}

\begin{abstract}
We have developed metal-oxide graphene field-effect transistors (MOGFETs) on sapphire substrates working at microwave frequencies.
For monolayers, we obtain a transit frequency up to $\sim$ 80~GHz
for a gate length of 200~nm, and a power gain maximum frequency of about $\sim$ 3~GHz for this specific sample.
Given the strongly reduced charge noise for nanostructures on sapphire, the high stability and high performance of this material
at low temperature, our MOGFETs on sapphire are well suited for a cryogenic broadband low-noise amplifier.

\end{abstract}


\maketitle

Graphene is a two-dimensional system with high carrier mobility that provides a rich playground for building novel high-frequency devices \cite{schwiertz2010}.
Although the characterization of graphene's electronic properties is one of the most active fields in condensed matter physics,
most experiments focus on its dc properties.

Metal-oxide graphene field-effect transistors (MOGFETs) operating in the microwave range have already been realized \,\cite{meric2008,lin2009,moon2009,lin2009a,farmer2009,lin2010a,lin2010,dimitrakopoulos2010,liao2010,lee2010,wu2011,meng2011}.
First microwave measurements on MOGFETs yielded transit frequencies of $f_{T} \sim 14.7$~GHz at 500-nm gate length\,\cite{meric2008}. These very promising results were later optimized further by reducing the gate length\,\cite{lin2009,lin2010a,lin2010,wu2011} and using different techniques to grow the top-gate oxide layer\,\cite{farmer2009}. Transit frequencies of 100~GHz have been reported for wafer-scale MOGFETs produced with graphene obtained by graphitization of SiC\,\cite{lin2010,dimitrakopoulos2010}. Values up to 155~GHz for a 40-nm gate length with an almost temperature independent gain\,\cite{wu2011} have been demonstated using MOGFETs fabricated with CVD-grown graphene\,\cite{lee2010,wu2011}.
Very recently transit frequencies up to 300~GHz have been reported using a Co$_2$Si nanowire as gate\,\cite{liao2010}.
Despite the lack of an energy gap making graphene unsuitable for logic applications, these results are very promising for the use of MOGFETs as microwave low-noise amplifiers. Still, graphene has yet to show its full potential and the "terahertz gap" remains to be bridged\,\cite{schwiertz2010}.

Reducing the gate length and the contact resistance, improving the carrier mobility and limiting parasitic capacitances should further improve the high-frequency performance of MOGFETs\,\cite{schwiertz2010,thiele2010,pince2010,liao2010a}.
Here, we present microwave measurements of MOGFET devices fabricated on sapphire substrates. We find transit frequencies up to 80~GHz for a gate length of 200~nm. The use of such a fully insulating substrate allows to minimize losses and parasitic capacitances arising from finite conduction of the substrate, which is present even in intrinsic semiconductors.
Charge noise is minimized on sapphire, which is therefore used as a substrate for microwave devices such as detectors and qubits. Moreover, because of the high energy of polar optical phonons in sapphire, the substrate-limited mobility for graphene is expected to be higher than on oxidized silicon.
Given the high thermal conductivity of sapphire, the very high cut-off frequency and the low impedance of the MOGFET, our device could be integrated in a monolithic microwave integrated circuit (MMIC) such as ultra-high bandwidth amplifiers.

\begin{figure}[htbp]
\scalebox{0.75}{\includegraphics{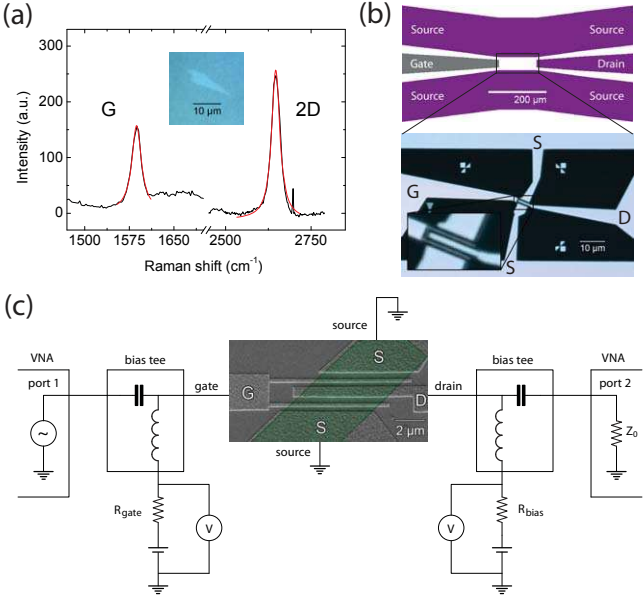}} {\caption{(a) Raman spectrum of the graphene monolayer on a C-plane sapphire substrate measured using a 1~mW laser excitation at a wavelength of 632.9~nm. The small variations in the background are due to the sapphire substrate. The G and 2D peaks are fitted with single Lorentzians. Inset: optical micrograph of the graphene sheet. (b) Schematics of the coplanar wave guide (top) and optical micrograph (bottom) of a measured MOGFET. (c) Schematics of the high-frequency measurement set-up.}}
\end{figure}

Graphene flakes were prepared using Scotch tape micromechanical cleavage on natural graphite, and deposited on 330 $\mu$m thick C-plane sapphire substrate. The graphene sheets were located using optical microscopy and characterized by Raman spectroscopy. Fig. 1(a) shows the Raman spectrum of the monolayer graphene flake on sapphire used for sample $SP15\_MD$. The samples were prepared by e-beam lithography. To avoid charging effects of the insulating substrate we used a layer of standard PMMA resist plus a second layer of water-soluble conducting polymer (Espacer 300Z from Showa Denko K.K.). The contacts were realized by  evaporating a Ti(10~nm)/Al(120~nm) bilayer. A 25-nm Al$_2$O$_3$ gate insulator was prepared by atomic layer deposition as described by Kim \emph{et al.}\,\cite{kim2009}. The dielectric constant of our oxide is estimated to $\kappa = 6.4$. A second lithography step allowed the patterning of the top gates followed by a Ti(10~nm)/Al(100~nm) bilayer deposition. We chose a geometry of a coplanar wave guide with double-gate structure as shown in Fig. 1(b). The measurements were performed at room and low temperatures (77~K) on RF probe stations. The voltage biases (both drain-source voltage $V_{ds}$ and gate voltage $V_g$) were applied using a voltage source and bias-tees (see Fig. 1(c)). The drain-source current $I_{ds}$ was deduced from the voltage drop across the bias resistor $R_{bias}$. The scattering parameters were measured with a two-port vector network analyzer (VNA Anritsu 37369C) calibrated using a short-open-load-thru calibration.
In order to extract the intrinsic transit frequency of the transistor, we applied a de-embedding procedure by subtracting the signal of an open and short structure strictly similar to the measured MOGFET using the relation: $Y_{de-emb}^{-1} = (Y_{fet}-Y_{open})^{-1}-(Y_{short}-Y_{open})^{-1}$, where $Y$ is the admittance calculated from the measured scattering parameters using conventional two-port network analysis\,\cite{pozar}. This technique has proven to be useful to probe diffusion in graphene\,\cite{pallecchi2011}. We will present measurements of device $SP15\_MD$, which showed the best performance in our batch of 10 samples. The double-gate device has a 200~nm gate length, 800~nm and 3.7~$\mu$m channel length $L$ and width $W$ respectively. From geometrical considerations, we estimate the total gate capacitance of the double-gate device $C_{g} \sim 3.5$~fF.

\begin{figure}[htbp]
\scalebox{0.4}{\includegraphics{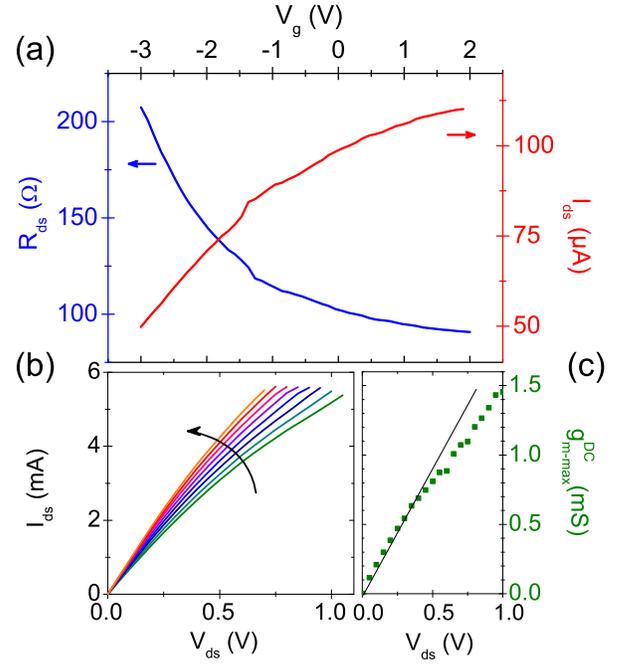}} {\caption{Measured dc characteristics of the MOGFET: (a) $R_{ds}$ and $I_{ds}$ versus $V_g$ measured at $V_{ds} = 10$~mV. (b) $I-V$ curves taken at $V_{g} = -2.0$~V to -0.25 V in 0.25-V steps (from bottom to top). (c) Maximum transconductance $g^{DC}_{m-max}$ versus $V_{ds}$, the black line is a guide for the eye.}}
\end{figure}

First we  characterize  the sample's dc properties to define the parameter range where the transconductance $g_m^{DC} = \frac{\partial I_{ds}}{\partial V_{g}}$ is maximum. Fig. 2(a) shows both the resistance and the drain-source current as a function of gate voltage $V_g$. The resistance was extracted from the $I-V$ curves at relatively low bias ($V_{ds}$ = 10~mV). Note the low impedance of the device which is of great advantage for integration in microwave circuits. We observed that in all the samples the graphene sheet is naturally n-doped, as expected for aluminum contacts. The Dirac point could not be reached within our gate-voltage range. Accordingly, the transconductance $g_m^{DC} (V_g)$ does not saturate but reaches a still considerable value $\frac{g_m}{2W\cdot V_{ds}}= 0.22$~mS/$\mu$mV.
Fig. 2(b) displays $I-V$ curves  of the MOGFET for several gate voltages becoming non-linear at large bias. We observe a tendency to saturate, but despite the large current passing through the device a full saturation is not found, as typically observed in graphene field-effect transistors\,\cite{meric2008a, bachtold}. Fig. 2(c) shows the maximum transconductance $g^{DC}_{m-max}$ versus $V_{ds}$, which deviates from linear behavior as the $I-V$ curves become non-ohmic.

\begin{figure*}[htbp]
\parbox{12cm}{\scalebox{0.35}{\includegraphics{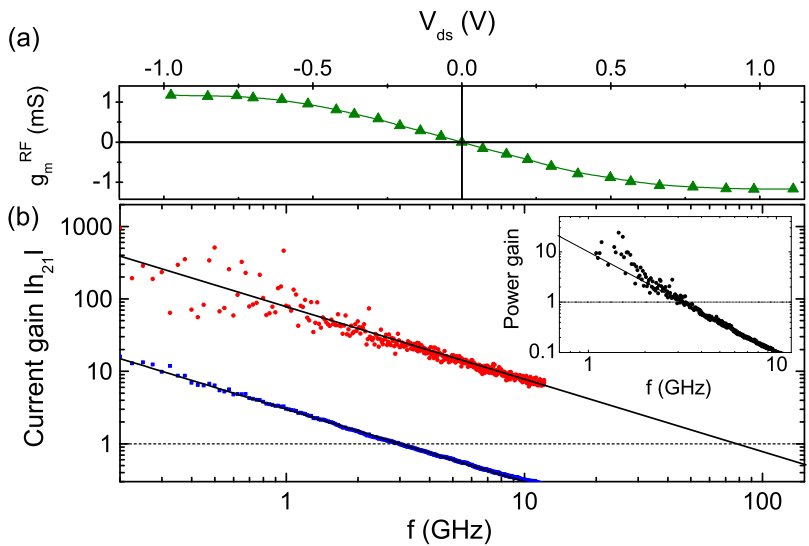}}}
\parbox{5cm}{\caption{Microwave frequency characterization of the $SP15\_MD$ MOGFET: (a) High-frequency transconductance $g_m^{RF}$ versus $V_{ds}$ for $V_{g}$ = -4.3~V. (b) Current gain $|h_{21}|$ as a function of frequency f for $V_{g}$ = -5.2~V and $V_{ds}$ = -1.1~V; The lower curve (small blue squares) corresponds to the raw data with a $f_T \sim$ 3~GHz and the upper curve (small red dots) corresponds to the de-embedded data with $f_T \sim$ 80~GHz. The inset shows the power gain as a function of frequency. For both current and power gain, a $1/f$ dependence is indicated as solid line.}}
\end{figure*}

We now discuss the RF properties of our MOGFET. The short-circuit current gain defined as 
$h_{21}(\omega) = \frac{Y_{21}}{Y_{11}} \approx j \frac{g_m}{\omega C_{g}}$
 is calculated from the scattering parameters $S_{ij}$ using the common-source small-signal equivalent circuit model\,\cite{pozar}. In Fig. 3(a), the high-frequency transconductance $g_m^{RF}$ corresponding to the real part of the admittance is displayed as a function of $V_{ds}$ for $V_g = -4.3$~V. As expected for a diffusive transistor the $g_m$ increases linearly with $V_{ds}$ at low bias and then starts to saturate.  We find a maximum transconductance $g_{m-max}^{RF} \sim$ 0.25~mS$/\mu$mV at $V_g = -5.2$~V and $V_{ds} = -1.1$~V. According to the definition of the current gain, the transit frequency  corresponds to the frequency at which the current gain is equal to 1 (see\,\cite{pozar} for example) and can be expressed as $f_T = \frac{g_m}{2 \pi C_{g}}$. We estimate the transit frequency of our device using the gate capacitance obtained from the geometry, $C_g = 3.5$~fF, and obtain a transit frequency $f_T \sim$ 70~GHz.

Fig. 3(b) shows the measured current gain $|h_{21}(\omega)|$ as a function of frequency obtained both from the raw data and after de-embedding.
In both cases a clear $1/f$ behavior is observed. The measured transit frequency of the raw data reaches $\sim$ 3~GHz while after de-embedding, we obtained $f_T \sim$ 80~GHz which is very close to the estimated value, confirming the accuracy of our de-embedding procedure. It is important to note the very high transit frequency despite the relatively low mobility of our top-gated devices ($200 \leq \mu \leq$ 500~cm$^{2}$s$^{-1}$V$^{-1}$).
The inset of Fig. 3(b) shows the two-port power gain\,\cite{pozar} calculated from the raw scattering parameters. We observe a maximum oscillation frequency $f_{max}$
of about $\sim$ 3~GHz that could be increased by optimizing the design of our coplanar wave guide. However, the small value of the ratio $\frac{f_{max}}{f_T}$ is
characteristic of graphene samples\,\cite{schwiertz2010}.
We point out that a high current bias induces additional doping of the graphene sheet leading to a shift of the Dirac point towards higher negative $V_g$. As a consequence, the highest value of $g_m$ may lie outside the accessible gate voltage range. Therefore, the full potential of the MOGFET may be even superior to what is shown by these measurements.

The sample was measured several times, and despite a long time between the measurements (up to two months), the only difference was a higher n-doping level and a lower impedance, but the microwave data were identical. Measurements at low temperature confirmed that the performance of the MOGFET is extremely stable, as also reported by Wu \emph{et al.}\,\cite{wu2011}, and remains strictly similar to the room-temperature data. We have also investigated a MOGFET based on bilayer graphene which showed much lower performance\,\cite{bilayer}.

To conclude, we have performed microwave measurements of MOGFETs on sapphire substrates. We find a large cut-off frequency of $\sim$ 80~GHz, which is only a factor of two smaller than state of the art MOGFET transistors with a 40-nm gate length, demonstrating the high performance of our MOGFET. Our results confirm the high potential of sapphire as a substrate for high-frequency ultra low noise graphene transistors. The performance of the MOGFET could be further improved by scaling down the gate length and by depositing graphene on boron nitride for example\,\cite{dean2010}. Low-temperature measurements demonstrate that our MOGFET on sapphire is compatible with the design of a broadband cryo-amplifier.

We acknowledge G. Dambrine, V. Derycke, H. Happy, O. Kr\"{o}mer and L. Petzold for fruitful discussions.
R.D.'s Shared Research Group SRG 1-33 received financial support by the Karlsruhe Institute of Technology within the framework of the German Excellence Initiative.
This work was supported by the EU project MMM@HPC FP7-261594 and the French contract No. ANR-2010-
BLAN-0304-01-MIGRAQUEL.

\end{document}